\documentclass[conference]{IEEEtran}
\IEEEoverridecommandlockouts
\usepackage{cite}
\usepackage{amsmath,amssymb,amsfonts}
\usepackage{algorithmic}
\usepackage{graphicx}
\usepackage{textcomp}
\usepackage{xcolor}
\usepackage{booktabs}   
\usepackage{array}      
\usepackage{subfig}

\usepackage{float}      

\usepackage{amsmath}

\def\BibTeX{{\rm B\kern-.05em{\sc i\kern-.025em b}\kern-.08em
    T\kern-.1667em\lower.7ex\hbox{E}\kern-.125emX}}
\begin{document}

\title{Robust Clustering Analysis of Genes Related to Age-related Macular Degeneration using RNA-Seq
}

\author{
\IEEEauthorblockN{Brayan Gutierrez}
\IEEEauthorblockA{\textit{Department of Statistics} \\
\textit{Rice University}\\
Houston, United States\\
bag5@rice.edu}
\and
\IEEEauthorblockN{Rinki Ratnapriya}
\IEEEauthorblockA{\textit{Department of Ophthalmology}\\
\textit{Baylor College of Medicine}\\
Houston, United States\\
rpriya@bcm.edu}
\and
\IEEEauthorblockN{Arko Barman}
\IEEEauthorblockA{\textit{Data to Knowledge Lab}\\
\textit{Rice University}\\
Houston, United States\\
arko.barman@rice.edu}

}


\maketitle

\begin{abstract}
Identifying genes associated with diseases is crucial to understanding disease mechanisms and developing therapies. However, identification of individual genes associated with a disease often needs to be supplemented with clustering analysis to understand the relationships between genes and identify gene modules beyond individual gene-level relationships. Gene co-expression networks are widely used as a graph theoretic approach to the clustering analysis of genes. In our work, we perform robust clustering analysis on RNA-Seq data of Age-related Macular Degeneration (AMD) patients and controls by generalizing one such framework, Multiscale Embedded Gene
Co-Expression Network Analysis (MEGENA). We propose a carefully curated set of module quality evaluation metrics to choose appropriate statistical distance-based or information theoretic similarity measures over simple linear correlation to represent the similarities between genes. Furthermore, we design and implement a stability test to ensure the robustness of the detected hub genes in the presence of noise. Finally, we propose differential module eigengene analysis for a deeper understanding of upregulation and downregulation of each module with respect to the disease and control groups for a comprehensive understanding of the clustering analysis. Besides detecting robust hub genes and modules that are supported by prior findings, we also identify previously undiscovered hub genes that can potentially lead to further biomedical research into understanding the AMD disease mechanism and developing new treatments.
\end{abstract}

\begin{IEEEkeywords}
RNA-Seq, co-expression networks, clustering, hub genes, transcriptomics
\end{IEEEkeywords}

\section{Introduction}

Age-related macular degeneration (AMD) is a progressive neurodegenerative disease that damages the macula, the retinal structure responsible for sharp central vision, color perception, and fine-detail recognition. AMD is the leading cause of irreversible vision loss in adults over the age of 50, affecting approximately 20 million individuals in the United States alone, with prevalence expected to increase substantially as the population ages. The disease manifests in two primary forms: dry (atrophic) AMD and wet (neovascular) AMD. Dry AMD, the more common form, is characterized by gradual thinning and degeneration of the macula and progresses through early, intermediate, and late stages, typically advancing slowly over time. Wet AMD is less common but more severe, and leads to accelerated vision loss. Importantly, late-stage dry AMD can progress further into wet AMD, underscoring the continuum of disease severity~\cite{ratnapriya_retinal_2019, ratnapriya_age-related_2013}.

Furthermore, AMD is a complex, multifactorial disease shaped by the cumulative effects of aging, genetic predisposition, and environmental factors~\cite{gorman_genome-wide_2024, fritsche_age-related_2014}. Despite extensive characterization of these risk factors, the molecular mechanisms underlying AMD progression and cellular vulnerability are incompletely understood~\cite{ma_integrating_2025}. However, in recent years, transcriptomics has emerged as a critical tool to identify genes associated with AMD~\cite{ash_role_2023, ratnapriya_retinal_2019}.

Comparative transcriptomic studies in disease-relevant tissues offer a powerful tool to uncover molecular mechanisms underlying AMD. However, such analyses face substantial challenges, including limited sample availability, biological heterogeneity across donors, and the high dimensionality of gene expression data, all of which increase susceptibility to false positives and reduce reproducibility when using traditional statistical approaches~\cite{patock_graphical_2025}. These challenges have motivated the increasing use of machine learning (ML) approaches, which are well-suited to modeling complex interdependencies in large-scale gene expression datasets. Recent studies have demonstrated improved robustness in identifying genes related to AMD~\cite{ma_integrating_2025, patock_graphical_2025}. Such approaches enable the discovery of candidate genes that may contribute to disease risk or progression, but do not explicitly capture the coordinated structure of gene regulation underlying complex phenotypes. Since complex diseases such as AMD arise from coordinated dysregulation of gene networks rather than isolated gene effects, network-based approaches have become essential complements to understand gene clusters and disease mechanisms using gene co-expression networks~\cite{tzec-interian_bioinformatics_2025, li_transcriptome_2022, morabito_hdwgcna_2023, ratnapriya_retinal_2019}. 

Gene co-expression network analysis is widely used in transcriptomics to identify groups, clusters, modules, or communities of genes that exhibit similar expression patterns in different cohorts of subjects~\cite{ratnapriya_retinal_2019, morabito_hdwgcna_2023, montenegro_gene_2022}. A co-expression network is a graph, $\mathcal{G} = (\mathcal{V}, \mathcal{E})$, where the nodes, $\mathcal{V}$, represent the genes and the edge weights, $\mathcal{E}$, represent the relationships between the genes. Most co-expression network construction strategies aim to make the network sparse rather than fully connected. As a result, the adjacency matrix of a co-expression network, $\mathcal{A}$, is a sparse matrix.

Within co-expression modules, not all genes contribute equally to network structure or function. Thus, co-expression network analysis often focuses on identifying hub genes. Hub genes are defined as genes that occupy central positions within modules, characterized by high connectivity or centrality relative to other genes. From a systems biology perspective, hub genes are of particular interest because they may act as key regulators, signal integrators, or bottlenecks that coordinate the behavior of multiple downstream genes~\cite{gui_identification_2021, bi_gene_2015, deng_predicting_2016}.


Most widely-used approaches for constructing and analyzing gene co-expression networks, such as Weighted Gene Co-expression Network Analysis (WGCNA)~\cite{langfelder_wgcna_2008}, high-dimensional WGCNA~\cite{morabito_hdwgcna_2023}, and Multiscale Embedded Gene Co-Expression Network Analysis (MEGENA)~\cite{song_multiscale_2015}, assume only linear relationships between gene expressions and utilize only Pearson correlation. In our methods, we attempt to generalize this process by considering statistical distance metrics and information theoretic metrics that capture more nuanced information between the expression values of genes by treating gene expressions as random variables. Furthermore, we design a stability test to understand the effect of noise and ensure that our results and findings are robust and reliable.

In this work, we identify robust hub genes for AMD using RNA-Seq data by generalizing MEGENA, a network-based framework designed to identify hierarchical, multiscale gene modules and associated hub genes from high-dimensional RNA-Seq data~\cite{song_multiscale_2015, song_building_2012}. MEGENA constructs a planar filtered network (PFN) to preserve biologically meaningful relationships while reducing noise, and then recursively identifies modules and submodules that represent coordinated gene programs across multiple scales. This multiscale perspective is particularly well-suited for AMD, where dysregulated immune, metabolic, and neurovascular pathways interact across biological scales. To generalize the MEGENA framework, we identified the best similarity metrics for gene-gene comparisons while constructing a PFN using an extensive set of module quality evaluation metrics. Furthermore, we designed and performed tests to ensure the stability and robustness of our findings. In addition, we conducted Differential Module Eigengene (DME) analysis~\cite{morabito_hdwgcna_2023} to identify modules that are upregulated or downregulated in late AMD and control groups. 


Our contributions can be summarized as follows:
\begin{itemize}
    \item We identify the best similarity metric between genes for the construction of a Planar Filtered Network (PFN) using a curated set of module quality evaluation metrics.
    \item We generalize Multiscale Embedded Gene Co-Expression Network Analysis (MEGENA) by leveraging statistical and information theoretic metrics instead of simple linear correlation to identify hierarchical, multiscale gene modules and associated hub genes related to AMD.
    \item We design and perform stability tests to ensure that the process of identifying hub genes is robust.
    \item We perform Differential Module Eigengene (DME) analysis on the gene modules to analyze which of the identified modules are upregulated or downregulated in the late AMD and control groups of subjects. 
\end{itemize}

\section{Related Work}


Advances in RNA-Seq technologies have enabled the generation of large-scale gene expression datasets across diverse tissues, conditions, and disease states. These data have been instrumental in understanding molecular mechanisms underlying complex diseases~\cite{tzec-interian_bioinformatics_2025, ma_integrating_2025, su_cell-type-specific_2023}. Network-based approaches have become central to transcriptomic studies, enabling the identification of coordinated gene programs and higher-order regulatory structure. Weighted Gene Co-expression Network Analysis (WGCNA) is one of the most widely used frameworks for constructing gene co-expression networks and identifying modules of highly correlated genes~\cite{langfelder_wgcna_2008}. Eigengene-based representations of modules further allow researchers to study relationships among modules and associate network structure with phenotypic traits~\cite{langfelder_eigengene_2007}. These approaches have been successfully applied across a range of diseases, including neurological, psychiatric, and oncological conditions.


Network and clustering approaches applied to RNA-seq data have demonstrated their utility in identifying disease-relevant gene signatures beyond traditional differential expression analyses. For example, co-expression and network-based methods have been used to distinguish glioblastoma from normal tissue using stem cell marker gene networks~\cite{mukherjee_quiescent_2020}, identify shared molecular pathways between sleep disorders and Alzheimer’s disease~\cite{liang_identification_2022}, and uncover common genetic architectures across suicidal ideation and suicide using brain and blood transcriptomes~\cite{sun_brain_2024}. These studies highlight the value of modeling gene expression as interconnected systems rather than isolated features.

Recent work has also explored increasingly sophisticated strategies for constructing and analyzing gene networks. Random walk–based methods and graph-theoretic techniques have been widely applied for learning structure in biological networks and propagating functional information across nodes~\cite{patock_graphical_2025}. Additionally, cell-type-specific co-expression inference methods using single-cell RNA-seq data have enabled more refined characterization of gene regulation within heterogeneous tissues~\cite{su_cell-type-specific_2023}. Multiscale Embedded Gene Co-Expression Network Analysis (MEGENA) was introduced to address limitations of single-scale network methods by explicitly capturing hierarchical and multiscale modular organization in gene expression data~\cite{song_multiscale_2015}. Together, these approaches emphasize the growing role of graph-based and multiscale representations in transcriptomic analysis.




\begin{figure*}[!t]
    \centering
    \includegraphics[width=\textwidth]{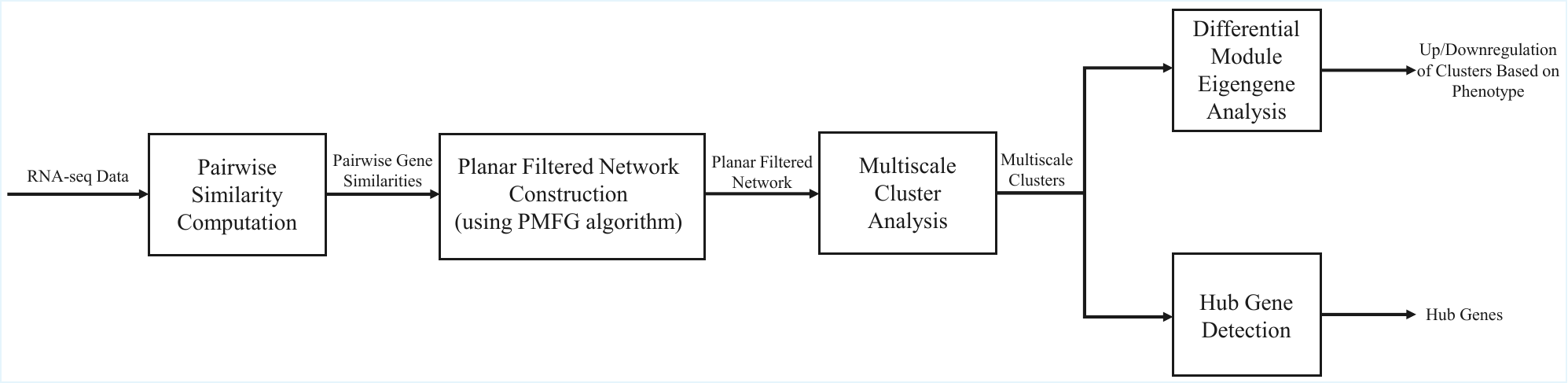}
    \caption{Block diagram of the RNA-seq network analysis workflow, from pairwise gene similarity computation and PFN construction to multiscale clustering, eigengene analysis, phenotype-specific regulation, and hub gene detection.}
    \label{fig:bd}
\end{figure*}

Research on AMD has shown complex disease traits, highlighting the interplay between genetic susceptibility, aging, and environmental factors~\cite{swaroop_genetic_2007}. Subsequent large-scale meta-analyses have identified the role of immune regulation, complement activation, lipid metabolism, and extracellular matrix remodeling in disease pathogenesis~\cite{han_genome-wide_2020}. As the majority of genetic risk factors reside in the non-coding regions of the genome, transcriptomic analyses have played a critical role in linking genetic risk variants to functional consequences in retinal tissues. Retinal RNA-seq and expression quantitative trait loci (eQTL) studies have identified genes whose expression is influenced by AMD-associated variants, providing mechanistic insight into how non-coding genetic variation contributes to disease risk~\cite{ratnapriya_retinal_2019, ash_role_2023}. Transcriptome-wide association studies spanning multiple tissues have further expanded the set of candidate genes potentially involved in AMD pathology, though interpretation remains challenging due to tissue specificity and regulatory complexity~\cite{strunz_transcriptome-wide_2020}. Comprehensive reviews of AMD genetics and transcriptomics emphasize that, despite substantial progress, a large fraction of disease heritability remains unexplained and that many implicated genes likely act through coordinated pathways rather than independent effects~\cite{bhumika_genetic_2024}. 

A recent study integrates explainable ML with bulk RNA-seq data to identify immune signatures associated with late-stage AMD. By applying ML-based feature selection to a large cohort of control and AMD donors, this study identified a curated set of candidate genes with strong discriminatory power. Importantly, these genes were analyzed using pathway enrichment and co-expression networks, highlighting the potential of combining ML-driven gene prioritization with network-based interpretation~\cite{ma_integrating_2025}. Complementary computational approaches have also been proposed to cluster AMD-associated genes using graph-based methods applied to co-expression networks, emphasizing robustness and stability through joint optimization of network construction, embedding, and clustering steps~\cite{patock_graphical_2025}. 



\section{Methods}

\subsection{Dataset}

Bulk RNA-Seq data includes transcriptomic profiles from human retinal tissue collected from control donors and individuals with late-stage AMD~\cite{ratnapriya_retinal_2019}. In our analysis, we use the data from $n_c=105$ controls and $n_s=61$ subjects with late-stage AMD. For our core analysis, we analyzed a curated set of 81 genes identified as highly informative for distinguishing control and late-stage AMD samples~\cite{ma_integrating_2025}. These genes were selected using explainable ML and statistical feature selection methods, thus representing biologically meaningful candidates enriched for immune, complement, and inflammatory pathways. Focusing on this gene set reduces dimensionality, mitigates noise associated with genome-wide analyses, and enables more interpretable network inference. We perform further robustness and stability testing to ensure the reliability of our results using samples from the complete RNA-Seq data~\cite{ratnapriya_retinal_2019}.

\subsection{Overview of our Analysis}

Fig.~\ref{fig:bd} shows an overview of our RNA-seq network analysis workflow used throughout this study, highlighting the key stages from raw expression data to hub gene identification and determination of upregulation and downregulation of clusters. In our work, we generalize the MEGENA framework~\cite{song_multiscale_2015} by including a comprehensive analysis of different similarity computation methods based on several module quality evaluation metrics, allowing a methodical and informed choice of similarity metric for calculating similarity between gene expression levels. Further, our methods involve eigengene computation and DME to derive information on the upregulation/downregulation of the identified gene clusters~\cite{langfelder_eigengene_2007, morabito_hdwgcna_2023}. 

\subsection{Pairwise Similarity Computation for Network Construction}

Starting from normalized RNA-Seq data, we compute pairwise gene similarities to quantify the relationship between all gene pairs. These similarities form the basis for downstream network construction and multiscale clustering.

Gene co-expression analysis depends critically on the choice of similarity measure used to quantify relationships between a pair of genes. In this study, we systematically evaluated the effectiveness of using different statistical and information theoretic similarity metrics rather than relying on a simple linear correlation-based measure used in MEGENA. In specific, we compared linear correlation measures (Pearson, Spearman, and Kendall), correlation-based network representations such as the WGCNA adjacency matrix~\cite{langfelder_wgcna_2008}, distance-based measures for probability distributions (Bhattacharyya distance and Wasserstein distance), and information theoretic approaches (Adjusted Mutual Information (AMI) and Jensen-Shannon divergence (JSD)). This structured exploration of similarity metrics allowed us to generalize MEGENA by considering metrics beyond simple linear correlation, and to assess the robustness of the inferred network structure and downstream modules using several module quality evaluation metrics.

Based on our preliminary exploratory analysis, Pearson correlation, AMI, and JSD were selected for detailed comparison, as they represent distinct modeling assumptions, produced stable network constructions suitable for MEGENA, and show better results in terms of the module quality evaluation metrics used in our study (see Section~\ref{subsec:module_qual_eval_metrics}). Pearson correlation provides a widely used linear baseline, AMI captures general nonlinear dependencies while correcting for chance associations, and JSD quantifies distributional similarity between gene expression profiles. For our dataset, other evaluated measures, including Spearman, Kendall, WGCNA-based adjacency, Bhattacharyya distance, and Wasserstein distance, exhibited greater sensitivity to noise, yielded worse results for the module quality evaluation metrics, or generated less consistent module structure across cohorts, and were therefore not pursued further in downstream analyses. We note that the choice of similarity metrics may be dependent on the dataset and encourage the exploration of all possible metrics mentioned here for other datasets. 

Our shortlisted similarity measures are described as follows:
\begin{itemize}

\item Pearson correlation provides a linear measure of association between two gene expression profiles
\(X=(X_1,\dots,X_n)\) and \(Y=(Y_1,\dots,Y_n)\), and is defined as,
\begin{equation}
\rho_{X,Y}
=
\frac{\sum_{i=1}^n (X_i-\bar X)(Y_i-\bar Y)}
{\sqrt{\sum_{i=1}^n (X_i-\bar X)^2}\sqrt{\sum_{i=1}^n (Y_i-\bar Y)^2}}.
\end{equation}
Pearson correlation captures linear dependence and is not suitable for capturing nonlinear associations and higher-order distributional differences.

\item Adjusted Mutual Information (AMI) is an information-theoretic similarity measure derived from mutual information (MI). For two discrete random variables \(X\) and \(Y\) with joint distribution \(p(x,y)\) and marginal distributions \(p(x)\) and \(p(y)\), mutual information is defined as
\begin{equation}
\mathrm{MI}(X,Y)
=
\sum_{x,y} p(x,y)\log\frac{p(x,y)}{p(x)p(y)}.
\end{equation}
To make MI more robust to noise, we calculate AMI by first performing quantile binning of the expression values for every gene, using $n_{bins}=5$ bins. Then, we calculate AMI as,
\begin{equation}
\mathrm{AMI}(X,Y)
=
\frac{\mathrm{MI}(X,Y)-\mathbb{E}[\mathrm{MI}(X,Y)]}
{\max\{H(X),H(Y)\}-\mathbb{E}[\mathrm{MI}(X,Y)]}
\end{equation}
where \(H(\cdot)\) denotes Shannon entropy and is defined by,
\begin{equation}
    H(X) = -\sum_{x \in \mathcal{X}} p(x) \log p(x)
\end{equation}
Since MI is biased upward for finite sample sizes, AMI corrects for chance agreement by normalizing MI with respect to its expected value under random labeling. Thus, this adjustment improves robustness and comparability across gene pairs, particularly in high-dimensional transcriptomic data where spurious associations are common~\cite{vinh_information_2010}.

\item Jensen--Shannon divergence (JSD) is a symmetric and bounded measure of dissimilarity between two probability distributions. Given two distributions \(P\) and \(Q\), JSD is defined as
\begin{equation}
\begin{aligned}
\mathrm{JSD}(P\|Q)
&=
\frac{1}{2}D_{KL}(P\|M)
+
\frac{1}{2}D_{KL}(Q\|M), \\
M
&=
\frac{1}{2}(P+Q).
\end{aligned}
\end{equation}
where $D_{KL}(\cdot\|\cdot)$ denotes the Kullback--Leibler divergence, and is given by,
\begin{equation}
    D(P\|Q) = \sum_{x \in \mathcal{X}} P(x) \log \frac{P(x)}{Q(x)}
\end{equation}
Unlike KL divergence, JSD is symmetric, bounded, and its square root defines a true metric on the space of probability distributions~\cite{endres_new_2003}. When applied to gene expression data, JSD compares the distribution of expression profiles, making it sensitive to changes in variability and multimodality that may not be reflected in correlation-based measures~\cite{lin_divergence_1991}.

\end{itemize}

\subsection{Planar Filtered Network (PFN) Construction}

Following similarity computation, the resulting weighted gene–gene relationships are used to construct a planar filtered network (PFN), a graph that can be drawn on the surface of a sphere with no intersecting edges~\cite{tumminello_tool_2005}. The PFN is constructed by leveraging the Fast Planar Filtered Network Construction (FPFNC) method~\cite{song_multiscale_2015}, an efficient modification of the Planar Maximally Filtered Graph (PMFG) algorithm~\cite{tumminello_tool_2005} that controls the false discovery rate (FDR) of similarity for each pair of genes. PFN construction preserves the most informative edges, enforcing sparsity while retaining the strongest associations, thus improving interpretability and computational tractability. 

\subsection{Multiscale Cluster Analysis}

Gene co-expression network analysis is motivated by the observation that genes involved in shared biological processes tend to exhibit coordinated expression patterns across samples. Rather than acting independently, genes often function as components of regulatory programs or pathways, and disruptions to these programs can manifest as disease phenotypes. Network-based approaches model these dependencies explicitly by representing genes as nodes and pairwise similarities in expression as weighted edges, enabling the identification of groups of tightly connected genes, commonly referred to as modules. Traditional co-expression frameworks such as WGCNA identify modules at a single scale by clustering genes based on correlation structure~\cite{langfelder_wgcna_2008}. While effective, single-scale methods may miss hierarchical organization, where large modules contain nested submodules corresponding to finer-grained biological processes.

Following PFN construction, we perform multiscale clustering analysis (MCA) to identify hierarchical gene modules or clusters~\cite{song_multiscale_2015}. MCA is a clustering procedure to identify modules and nested submodules across multiple scales, explicitly modeling hierarchical organization in gene regulation. It utilizes the following three unrelated criteria for the identification of locally coherent clusters while preserving a globally optimum partition:
\begin{itemize}
    \item Shortest Path Distances ($SPD$)~\cite{albert_statistical_2002} between nodes were used for the optimization of within-cluster compactness
    \item Local Path Index ($LPI$)~\cite{lu_similarity_2009} was utilized for optimizing the local clustering structure
    \item Overall modularity ($Q$)~\cite{newman_modularity_2006} was leveraged for the identification of optimal partition 
\end{itemize}

MCA leverages a measure of network compactness,
\begin{equation}
    \nu = \frac{\overline{SPD}}{\log |V|^\alpha}
\end{equation}
where $\overline{SPD}$ is the mean of shortest path distances for all node pairs, $V$ is the set of nodes or vertices in the graph, and $\alpha$ is a resolution parameter. Starting with the entire network as a single cluster, MCA iteratively performs a search through a range of values for $\alpha$ to identify resolution parameter values for child clusters that are more compact than a parent cluster for multiscale clustering. 

Finally, module eigengenes are computed and tested for differential regulation across phenotypes, enabling the identification of up- and down-regulated clusters. Hub gene detection is performed within significant modules to highlight highly connected and potentially biologically influential genes. Together, this workflow provides a unified framework for examining how alternative pairwise similarity definitions propagate through network topology, module structure, and hub gene discovery.

\subsection{Hub Gene Identification}


We performed Multiscale Hub Analysis (MHA)~\cite{song_multiscale_2015} to identify hub genes associated with AMD from our data. MHA detects the nodes that are highly connected with other nodes in the multiscale clustering results using the within-cluster connectivity of nodes. The within-cluster connectivity of a node, $v_i$ at scale $\alpha$ defined as,
\begin{equation}
    c^w (v_i, \alpha) = \sum_{v_j \in V_{\ell}^{\alpha}} \mathcal{A}(v_i, v_j)
\end{equation}

where $V_{\ell}^{\alpha}$ is the set of nodes in cluster $\ell$ at scale $\alpha$, and $\mathcal{A}$ is the adjacency matrix.

\subsection{Differential Module Eigengene Analysis}

While gene-level analyses provide valuable insight, interpreting disease-associated changes at the level of entire modules requires a reduced representation of module activity. Eigengenes were introduced by~\cite{langfelder_eigengene_2007} as a principled way to summarize the expression profile of a gene module using a single quantitative variable. Specifically, a module eigengene is defined as the first principal component of the standardized expression matrix of genes within a module, capturing the dominant pattern of variation shared across those genes. Eigengenes offer several advantages for downstream analysis. First, they reduce dimensionality, allowing complex gene sets to be represented by a single variable that can be directly associated with phenotypic traits. Second, they are robust to noise at the individual gene level, as they reflect coordinated expression rather than isolated fluctuations. Third, eigengenes facilitate statistical testing and visualization of module-level differences across conditions, such as disease versus control~\cite{langfelder_eigengene_2007}.

Differential Module Eigengene (DME) analysis offers a mechanism to link co-expression modules to external traits, including clinical phenotypes, disease status, and environmental exposures~\cite{langfelder_eigengene_2007, morabito_hdwgcna_2023}. In the context of AMD, eigengene-based analyses enable direct testing of the upregulation and downregulation of individual modules or clusters systematically between control and late-stage disease. 


\section{Experiments \& Results}

\begin{table}[t]
\centering
\caption{Comparison of module quality evaluation metrics for Pearson correlation, AMI, and JSD.}
\label{tab:metric_comparison}
\footnotesize
\setlength{\tabcolsep}{4pt}
\begin{tabular}{lccccc}
\toprule
\textbf{Metric} 
& \textbf{Modularity} 
& \begin{tabular}{@{}c@{}}\textbf{Mean} \\ \textbf{Conductance}\end{tabular} 
& \begin{tabular}{@{}c@{}}\textbf{Mean} \\ \textbf{Density}\end{tabular} 
& \begin{tabular}{@{}c@{}}\textbf{Mean} \\ \textbf{Transitivity}\end{tabular} 
& \begin{tabular}{@{}c@{}}\textbf{Mean} \\ \textbf{Correlation}\end{tabular} \\
\midrule
Pearson 
& 0.5056 
& 0.2394 
& 0.2373 
& 0.4187 
& 0.6444 \\
AMI 
& 0.5438 
& 0.2088 
& 0.3254 
& 0.4810 
& 0.6373 \\
JSD 
& 0.3260 
& 0.0953 
& 0.2215 
& 0.4122 
& 0.6533 \\
\bottomrule
\end{tabular}
\end{table}

\subsection{Module Quality Evaluation Metrics}
\label{subsec:module_qual_eval_metrics}
To quantitatively assess the different similarity metrics, we evaluated five complementary graph-based metrics that characterize distinct aspects of modular organization. Together, these metrics provide a robust assessment of module coherence beyond modularity alone and are commonly used in gene co-expression network analysis~\cite{song_multiscale_2015,langfelder_wgcna_2008}. Our chosen module quality evaluation metrics are:

\begin{itemize}

\item Modularity is a scalar measure that quantifies the strength of community structure in a network by comparing the density of edges within modules to the expected number of edges in the null model, i.e, a fully random network. For an undirected graph, modularity is defined as,
\begin{equation}
    Q = \frac{1}{2m}
    \sum_{i,j} \left( \mathcal{A}_{ij}  - \frac{k_i k_j}{2m} \right)
    \delta(c_i, c_j),
\end{equation}
where $\mathcal{A}_{ij}$ denotes the $(i,j)$-th element of the adjacency matrix, $k_i$ and $k_j$ are the degrees of nodes $i$ and $j$ respectively, $m$ is the total number of edges in the network, $c_i$ is the community assignment of node $i$, and $\delta(c_i, c_j)$ is the Kronecker delta function. Higher modularity indicates stronger separation between modules relative to random expectation, i.e., a high density of edges within modules and sparse edges between modules~\cite{newman_modularity_2006}.
    
\item Conductance measures the extent to which a module is separated from the remainder of the network, with lower values indicating stronger isolation and reduced edge leakage.~\cite{yang_defining_2012} It is defined as,

\begin{equation}
\phi(M)=
\frac{
|\{(i,j): i\in M,\ j\notin M\}|
}
{\sum_{i\in M} \deg(i)}.
\end{equation}

\item Density quantifies the fraction of possible edges that are present within a module, providing a measure of internal connectivity. Density is defined as,

\begin{equation}
\text{dens}(M)=
\frac{2|E_M|}{|M|(|M|-1)}.
\end{equation}

\item Transitivity (global clustering coefficient) measures the prevalence of closed triplets, reflecting the tendency of genes within a module to form tightly connected neighborhoods.

\begin{equation}
C =
\frac{\text{closed triplets}}
{\text{all triplets}}.
\end{equation}

\item Although correlation was not used as the primary edge weight in our AMI-based generalized MEGENA networks, mean within-module correlation for all modules provides a useful diagnostic of co-expression coherence. It is given by,

\begin{equation}
\text{AvgCorr}(M)=
\frac{1}{\binom{|M|}{2}}
\sum_{i<j} \text{cor}(x_i, x_j).
\end{equation}

\end{itemize}

Based on the results of these metrics for all our similarity metrics (Pearson correlation, AMI, and JSD), we primarily focused on AMI for further analysis (see Table~\ref{tab:metric_comparison}). While multiple similarity measures were explored, AMI provided the most favorable balance of module separation, internal coherence, and stability across cohorts, making it the most appropriate choice for hierarchical multiscale network inference in this study.

\subsection{Hyperparameter Selection}

MEGENA identifies hierarchical module structures in co-expression networks through PFN construction and recursive module detection~\cite{song_multiscale_2015}. Since network topology and module stability can be sensitive to algorithmic parameters, hyperparameters were selected based on recommendations from the original MEGENA framework and values commonly adopted in prior biomedical applications~\cite{liang_identification_2022, mukherjee_quiescent_2020, sun_brain_2024, ru_integrated_2025}. Specifically, module significance was assessed using a permutation-based test with a module-level significance threshold of $p_{mod} \leq 0.05$, such that only modules with empirical $p$-values below this cutoff were retained. Hub gene significance within modules was evaluated using an analogous permutation procedure with a hub-level threshold of $p_{hub} = 0.05$, ensuring that identified hubs exhibited connectivity patterns unlikely to arise by chance.

To avoid unstable or biologically uninterpretable small clusters, a minimum module size constraint of $n_{min}^{mod} = 10$ genes was imposed, consistent with prior MEGENA studies in high-dimensional transcriptomic data. Statistical significance for both module and hub identification was estimated using $n_{perm} = 100$ permutations, balancing computational efficiency with robustness of inference. These hyperparameter settings were fixed across all analyses to facilitate comparability across cohorts and robustness experiments. While alternative parameter configurations continue to be explored, these literature-aligned choices were adopted to reduce sensitivity to stochastic variation and improve reproducibility in multiscale gene co-expression networks.

\begin{figure}[!t]
    \centering
    \includegraphics[trim = 0 0 0 0, clip, width=0.495\textwidth]{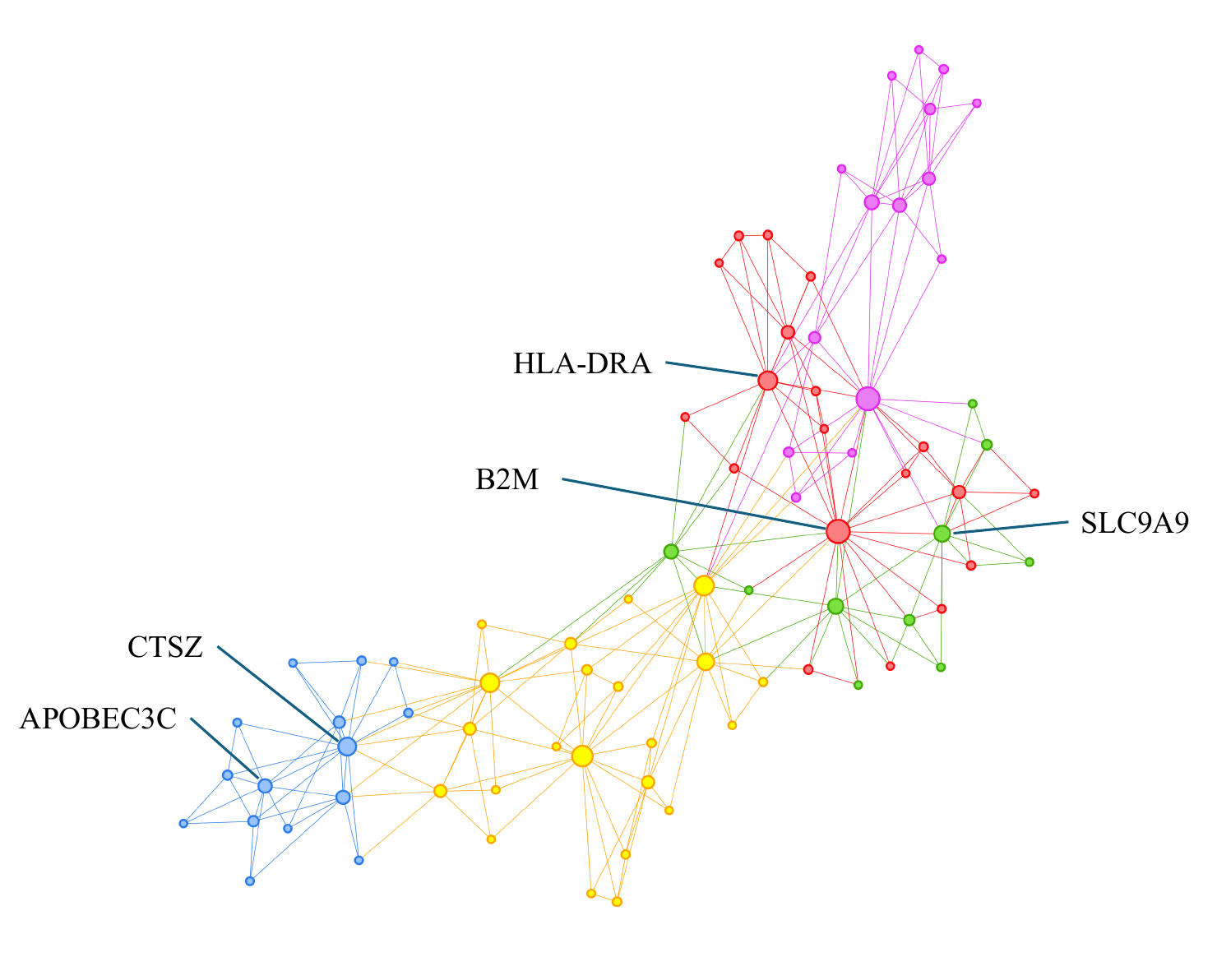}
    \caption{AMI-based generalized MEGENA network of the 81-gene AMD set for the \textit{combined cohort}, with nodes colored by multiscale module and key hub genes labeled, illustrating their centrality in the combined cohort.}
    \label{fig:ami_hub}
\end{figure}




\subsection{Results}


To understand and analyze the structure and identify robust hub genes, we applied our methods to the dataset in three different ways: (i) \textit{Combined cohort:} both control and late AMD groups; (ii) \textit{Control cohort:} only control group; (iii) \textit{Disease cohort:} only late AMD group.

\subsubsection{Module-Level Structure}

Our generalized MEGENA network, constructed with AMI as the similarity metric, revealed well-defined multiscale modular organization within the \textit{combined cohort} (Fig.~\ref{fig:ami_hub}), \textit{control cohort} (Fig.~\ref{fig:ami_100_1}), and \textit{disease cohort} (Fig.~\ref{fig:ami_100_4}). Detected modules exhibited high internal density and transitivity, along with reduced conductance, indicating strong separation from the remainder of the network. These properties suggest that AMI effectively captures coordinated transcriptional structure among the curated AMD-associated genes. 

\subsubsection{Hub Gene Identification}

Within AMI-based multiscale modules for the \textit{combined cohort}, we identified the following genes as hub genes across hierarchical levels: \textit{SLC9A9}, \textit{B2M}, \textit{HLA-DRA}, \textit{CTSZ}, and \textit{APOBEC3C}. The recurrence of these hubs across multiscale resolutions suggests that their centrality is not driven solely by network size, or gene set restriction, but instead reflects stable topological importance within the inferred network structure (Fig.~\ref{fig:ami_hub}). 

\begin{figure*}[!t]
    \centering
    \subfloat[AMI network structure for the \textit{control cohort}]{
        \includegraphics[
            width=0.48\textwidth,
            trim=0cm 0cm 0cm 0cm,
            clip
        ]{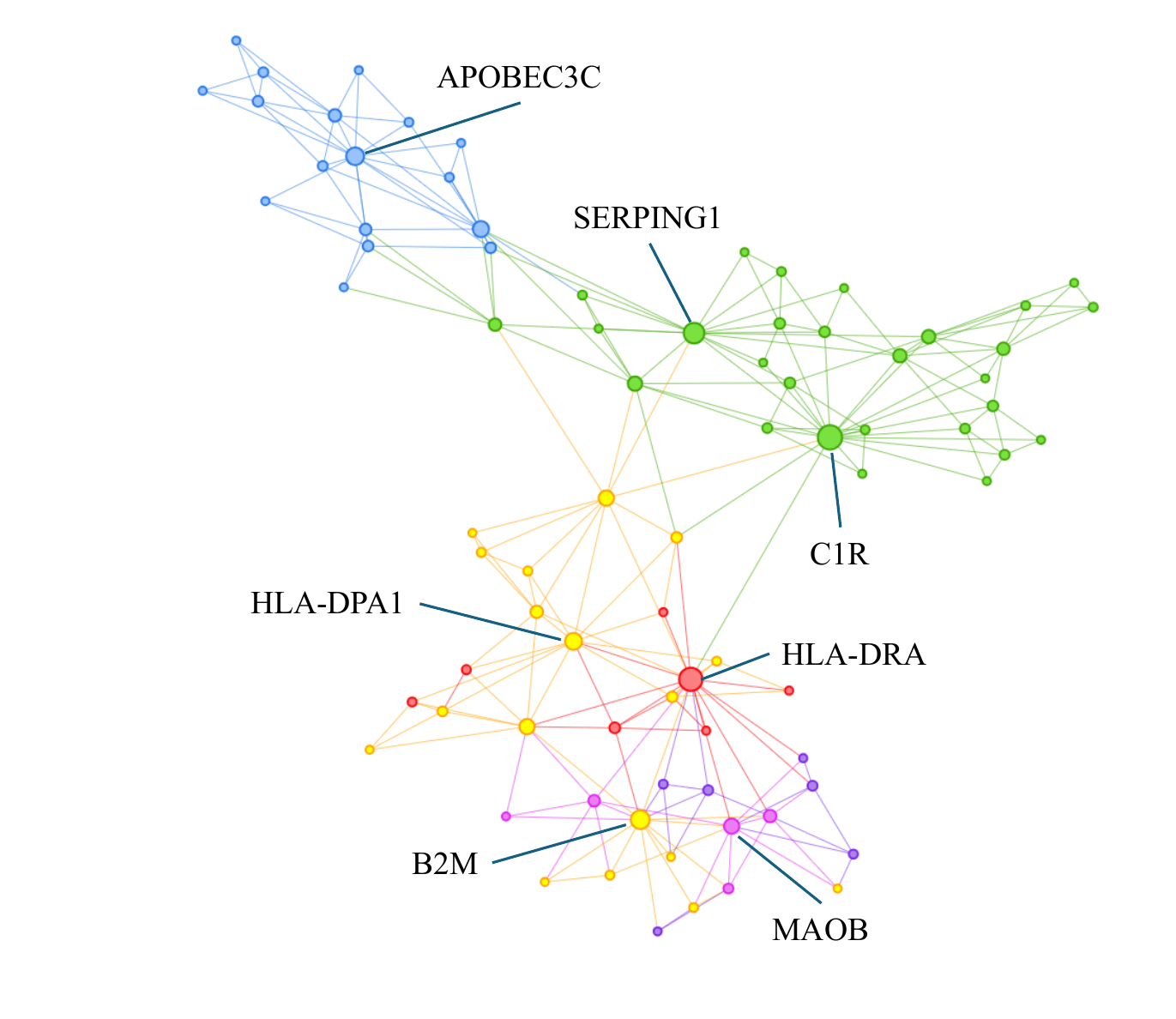}
        \label{fig:ami_100_1}
    }
    \hfill
    \subfloat[AMI network structure for the \textit{disease cohort} (late AMD)]{
        \includegraphics[
            width=0.48\textwidth,
            trim=0cm 0cm 0cm 0cm,
            clip
        ]{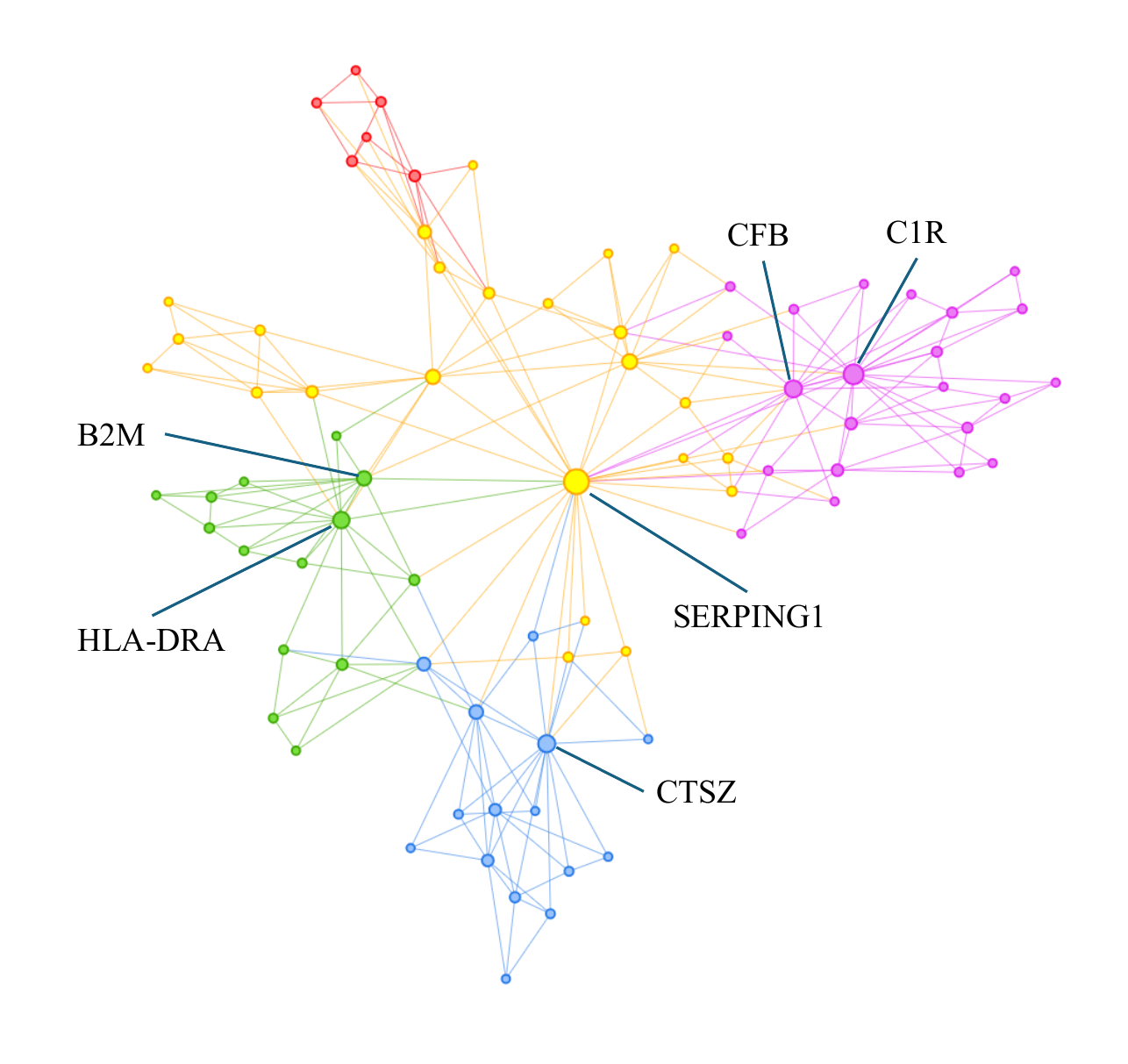}
        \label{fig:ami_100_4}
    }
    \caption{AMI-based MEGENA network of the 81-gene AMD
set, with nodes colored by multiscale module and key hub
genes labeled, illustrating their centrality in the control and late cohorts. Note the different graph structures for the two cohorts.}
    \label{fig:ami_100_compare}
\end{figure*}

In the \textit{control cohort}, \textit{C1R}, \textit{SERPING1}, \textit{HLA-DRA}, \textit{B2M}, \textit{APOBEC3C}, \textit{HLA-DPA1}, and \textit{MAOB} were identified as prominent hubs (Fig.~\ref{fig:ami_100_1}), whereas in the \textit{disease cohort}, we identified \textit{C1R}, \textit{CFB}, \textit{SERPING1}, \textit{HLA-DRA}, \textit{B2M}, and \textit{CTSZ} as the hub genes (Fig.~\ref{fig:ami_100_4}). 

Notably, \textit{C1R}, \textit{SERPING1}, \textit{HLA-DRA}, and \textit{B2M} were shared hub genes across both \textit{control} and \textit{disease} cohorts, suggesting that core immune- and complement-associated structure is preserved under AMI-based generalized MEGENA network construction. In contrast, cohort-specific hubs, such as \textit{APOBEC3C}, \textit{HLA-DPA1}, and \textit{MAOB} for the \textit{control cohort}; and \textit{CFB} and \textit{CTSZ} for the \textit{disease cohort} indicate additional network topology shifts that may reflect phenotype-dependent regulation. 

Furthermore, we note that \textit{HLA-DRA} and \textit{B2M} were identified as hub genes in all three of our networks. Overall, the recurrence of shared hubs across cohorts supports the stability of AMI-derived hub structure and suggests that the highest-centrality genes are not driven solely by network size or sampling variability, but instead reflect consistent topological importance within the inferred multiscale architecture (Fig.~\ref{fig:ami_100_compare}).

\subsection{Stability Testing for Robustness and Reliability}

We designed a complementary robustness test, where the set of 81 AMD-associated genes identified by statistical and ML methods~\cite{ma_integrating_2025} was appended with 419 randomly selected genes from the complete RNA-Seq data~\cite{ratnapriya_retinal_2019} to construct a generalized MEGENA network having $500$ genes and identify hub genes. This experiment was repeated $100$ times, and the number of times an originally identified hub gene was still identified as a hub gene was noted. This design directly tests whether the hub genes that we identified retain centrality in co-expression networks even in the presence of a large pool of randomly selected genes under identical network constraints.

Our generalized MEGENA networks demonstrated strong preservation of hub structure in this stability test (Table~\ref{tab:robust2_ami}). Core hub genes identified in the 81-gene set continued to appear with high frequency across repeated runs, indicating that their centrality reflects stable co-expression relationships rather than sensitivity to dataset composition or dimensionality. Notably, \textit{C1R}, \textit{HLA-DRA}, \textit{B2M}, and \textit{SERPING1} were among the most frequently detected hub genes across the cohorts individually as well as taken together. Collectively, these findings support the robustness of AMI-derived hub identification across multiple randomization strategies, while ongoing analyses further explore alternative sampling schemes and parameter settings.

In addition, this analysis aids in identifying only the most robust hub genes detected in the generalized MEGENA process while discarding the ones that are more sensitive to noise. The importance of this analysis lies in the fact that the RNA-Seq dataset used in our analyses is potentially noisy.

\begin{table}[t]
\centering
\caption{AMI hub gene frequencies for the stability experiment, where the original 81 AMD-associated genes were expanded with 419 different randomly sampled genes from the entire RNA-Seq data in each of the 100 repeated runs for a total of 500 for each cohort.}
\label{tab:robust2_ami}
\footnotesize
\begin{tabular}{lcccc}
\toprule
\textbf{Gene} & 
\begin{tabular}{@{}c@{}} \textit{Combined}\\ \textit{Cohort}\end{tabular} & 
\begin{tabular}{@{}c@{}} \textit{Control}\\ \textit{Cohort}\end{tabular} & 
\begin{tabular}{@{}c@{}} \textit{Disease}\\ \textit{Cohort}\end{tabular} \\
\midrule
C1R        & 99 & 94 & 93\\
HLA-DRA   & 84 & 70 & 81 \\
SERPING1  & 84 & 43 & 94 \\
B2M & 66 & 64 & 46 \\
CFB & - & - & 65 \\
CTSZ & 17 & - & 59\\
APOBEC3C & 56 & 39 & - \\
HLA-DPA1      & - & 21 & - \\
MAOB & - & 13 & - \\
SLC9A9        & 1 & - & -\\
\bottomrule
\end{tabular}
\end{table}

\subsection{Differential Module Eigengene Analysis}

Eigengenes for all modules were computed as the first principal component of standardized expression matrices for each module in the generalized MEGENA network of the \textit{combined cohort}~\cite{langfelder_eigengene_2007, morabito_hdwgcna_2023}. Eigengenes provide a compact summary of coordinated module activity and facilitate statistical association with disease status. Note that this analysis relies on the presence of two phenotypes for the detection of upregulation/downregulation of modules and can, thus, only be performed for the \textit{combined cohort}, which contains both \textit{disease} and \textit{control} cohorts. 

Our analyses indicate that several identified modules exhibit shifts in eigengene expression between \textit{control} and \textit{disease} (late-stage AMD) cohorts, consistent with coordinated transcriptional reprogramming. In particular, we noted that the module shown in green in Fig.~\ref{fig:ami_hub} exhibits downregulation when comparing the \textit{disease} with the \textit{control} samples. All of the four other modules (shown in red, magenta, yellow, and blue in Fig.~\ref{fig:ami_hub}) show upregulation when comparing the \textit{disease} with the \textit{control} samples.

To test the statistical significance of our results, we performed a t-test as well as a Wilcoxon rank-sum test for the DME analyses for each of the modules. Both tests yielded p-values of $p < 10^{-5}$ for every module, indicating that our findings for the DME analysis are statistically significant. The observed trends align with similar immune and glial transcriptional signatures previously reported in other studies~\cite{ma_integrating_2025}.

\section{Discussion}

\subsection{Key Findings and Biological Insights}

In this study, we generalized the MEGENA framework to incorporate an informed choice of gene-gene similarity metric among a wide variety of statistical distance metrics and information theoretic similarity measures instead of assuming simple linear relationships captured by Pearson correlation. We leveraged a curated set of module quality evaluation metrics to capture different desirable properties of modules to make this choice. In addition, we designed and performed a stability test to check the robustness of the detected hub genes in the presence of noise. We applied our methods to detect robust hub genes for the \textit{control}, \textit{disease}, and \textit{combined} cohorts in a dataset consisting of late-stage AMD patients and controls. Finally, we performed DME analysis to identify which of the detected modules in our \textit{combined cohort} are upregulated/downregulated when comparing the \textit{disease cohort} (late-stage AMD patients) with the \textit{control cohort}. Across experiments, our analyses consistently highlighted immune, complement, and inflammatory genes as central components of AMD-associated transcriptional organization, reinforcing the role of immune dysregulation as a core driver of disease progression, as seen in prior studies~\cite{ratnapriya_retinal_2019, ma_integrating_2025}.

Using AMI-based generalized MEGENA, we identified the following highly robust hub genes when considering the \textit{combined cohort}: \textit{HLA-DRA}, \textit{B2M}, and \textit{APOBEC3C}. These genes are well-established contributors to complement activation and antigen presentation pathways and have been repeatedly implicated in AMD genetics and retinal transcriptomics~\cite{ratnapriya_retinal_2019, ash_role_2023}. Their persistence as hubs even under substantial random background expansion suggests that they represent fundamental regulatory anchors of AMD-associated gene networks rather than artifacts of network size or gene selection.

On a similar note, we identified the following robust hub genes in the \textit{control cohort}: \textit{C1R}, \textit{HLA-DRA}, and \textit{B2M}; and the following robust hub genes in the \textit{disease cohort}: \textit{C1R}, \textit{SERPING1}, \textit{HLA-DRA}, \textit{CFB}, and \textit{CTSZ}. It is promising to note that while \textit{C1R} and \textit{HLA-DRA} were detected separately for both groups, there are several robust hub genes identified only in the \textit{disease cohort} that have not been previously identified in transcriptomics studies on AMD. 


At the module level, DME analysis revealed large and statistically significant shifts between \textit{control} and \textit{disease} (late-stage AMD) groups across all modules, indicating coordinated transcriptional reprogramming rather than isolated gene-level changes. Together, these findings support a systems-level view of AMD progression in which disease-associated pathways reorganize into more tightly coupled regulatory programs.

\subsection{Methodological Advances Over Prior Models}

Our framework advances prior transcriptomic network analyses in several important ways. First, rather than relying exclusively on simple linear correlation, we explored a large set of information-theoretic and statistical distance-based similarity measures (AMI, JSD, Pearson correlation, Spearman coefficient, Kendall coefficient, WGCNA adjacency matrix, Bhattacharyya distance, and Wasserstein distance) to construct co-expression networks. These measures capture nonlinear dependencies and distributional shifts in gene expression that are not detectable by correlation alone, addressing known limitations of traditional co-expression methods~\cite{langfelder_wgcna_2008,patock_graphical_2025}. While we selected AMI for our dataset, our methods indicate that an informed choice should be made based on the particular dataset being analyzed. For making this informed choice of gene-gene similarity measure, we propose a curated set of module quality evaluation metrics, including modularity, mean conductance of all modules, mean density of all modules, mean transitivity of all modules, and mean within-module correlation of all modules. These metrics capture a set of desired characteristics of modules, helping us in empirically choosing the most appropriate similarity measure. 

Second, by integrating these similarity measures with MEGENA, we explicitly modeled hierarchical and multiscale network structure. Unlike single-scale clustering approaches, MEGENA enables the detection of nested modules and submodules, which is particularly well-suited for complex diseases characterized by interacting biological processes~\cite{song_multiscale_2015}. This multiscale perspective allowed us to identify both broad immune-centered programs and finer-grained disease-specific regulatory structures.

Third, we incorporated explicit stability experiments to assess the robustness of the detected hub genes in the presence of other randomly selected genes. Many network studies report hubs identified from a fixed gene set without evaluating whether those hubs remain central in larger, more realistic transcriptomic contexts. Our robustness analyses demonstrate that several of our AMI-derived hub genes are highly stable under such dataset perturbations, providing stronger evidence that these genes represent biologically meaningful network regulators.

\subsection{Implications for AMD Research}

From a disease-specific perspective, our results complement and extend prior AMD transcriptomic studies. Previous work has successfully identified AMD-associated genes using GWAS, eQTL analyses, and explainable machine learning~\cite{han_genome-wide_2020,ratnapriya_retinal_2019,ma_integrating_2025}. However, these approaches primarily focus on studying individual genes rather than network organization.

By explicitly modeling co-expression structure among ML-prioritized genes, our analysis reveals how these genes interact as coordinated regulatory programs. The emergence of stronger modularity, increased transitivity, and higher within-module correlation in Late AMD networks suggests that disease progression is accompanied by increased regulatory coupling that could result in reaching a threshold for photoreceptor degeneration.


\subsection{Biological Validation and Therapeutic Potential}

The hub genes identified in this study represent strong candidates for further biological validation. Immune and complement hubs such as \textit{C1R}, \textit{SERPING1}, and \textit{CFB} are already implicated in AMD pathogenesis and may serve as biomarkers of disease state or progression~\cite{bradley_complement_2011, ennis_association_2008}. Disease-specific hubs such as \textit{CFB} and \textit{CTSZ}, which show increased centrality specifically in late-stage AMD, may represent novel regulatory nodes whose functional roles warrant experimental investigation.


From a translational perspective, network-central genes may offer advantages as therapeutic targets because perturbing a hub can influence entire regulatory programs rather than individual downstream genes. While direct gene therapy targeting immune hubs must be approached cautiously due to pleiotropic effects, network-based prioritization can help identify candidates for pathway-level modulation or combination therapies. These results provide a foundation for integrating transcriptomic network analysis with functional assays, cellular models, and in vivo validation~\cite{michael_pleiotropic_2025}.

\subsection{Limitations, Generalizability, and Future Directions}

Our study demonstrates that integrating information-theoretic similarity measures with multiscale network analysis provides a powerful and generalizable approach for uncovering coordinated transcriptional programs in complex diseases. In conjunction with stability tests to detect only the most robust hub genes and DME analysis for identifying upregulation/downregulation of modules, our methods can be applied to any RNA-Seq dataset for the robust identification of hub genes. In the context of AMD, our findings reinforce the central role of immune dysregulation while highlighting disease-emergent regulatory structures that may inform future mechanistic and therapeutic research.

While our methods show promising results, we note that our dataset includes a limited number of subjects ($n=166$). Although this sample size is not unusual for retinal transcriptomic studies, datasets from larger cohorts may lead to more statistically robust findings. Furthermore, we emphasize that our methods are designed and developed to be used with smaller curated gene subsets rather than entire RNA-seq datasets. In this work, we ensured the statistical stability and robustness of our findings through our experiment design for use in smaller datasets. The application of our methods to entire RNA-seq datasets will be carefully evaluated in our future work.

Although this study focuses on AMD, the proposed framework is broadly applicable to other diseases and transcriptomic datasets. The combination of ML-based gene prioritization, similarity measures based on statistical distance metrics and information theoretic similarity metrics, multiscale network analysis, robustness testing, and DME analysis can be readily applied to RNA-Seq studies of other neurodegenerative, autoimmune, or complex multifactorial diseases.

Our future work will extend this approach to genome-wide gene sets, integrate single-cell or spatial transcriptomic data, or incorporate additional layers of biological information such as epigenomic or proteomic data. Applying this framework longitudinally or across disease subtypes may further elucidate how network organization evolves over time or in response to treatment.

\section{Conclusion}

Our work presents a robust, network-based strategy for robust clustering analysis to identify stable hub genes and coordinated transcriptional programs associated with late-stage AMD using bulk RNA-seq data. Building on a set of genes identified by statistical and explainable ML methods~\cite{ma_integrating_2025}, we integrate information-theoretic and statistical distance-based similarity measures with multiscale embedded gene co-expression network analysis (MEGENA) to characterize disease-associated regulatory organization at both the module and hub levels~\cite{song_multiscale_2015}.

Across \textit{control}, \textit{disease} (late-stage AMD), and \textit{combined} cohorts, our results indicate that AMI-weighted generalized MEGENA networks can yield coherent multiscale modules and robust hub patterns that are robust to noise and show the usefulness of information theoretic similarity metrics over linear correlation. In particular, AMI-based generalized MEGENA networks consistently prioritize immune and complement-centered hub genes such as \textit{C1R}, \textit{SERPING1}, \textit{HLA-DRA}, \textit{B2M}, and \textit{CFB}, which remain aligned with established AMD genetics and immunopathology~\cite{ratnapriya_retinal_2019,bhumika_genetic_2024}. 


Our robustness analyses evaluate whether hub genes persist in the presence of noise introduced by randomly sampling a large number of genes from the complete RNA-Seq data. We show that several AMI-derived hub genes are comparatively stable across cohorts in our stability experiments. At the module level, eigengene-based DME analyses show coordinated transcriptional shifts between phenotypes, providing a module-centric complement to gene-level prioritization~\cite{langfelder_eigengene_2007}. 




\bibliographystyle{IEEEtran}
\bibliography{references}

\end{document}